\documentclass{article}

\usepackage{arxiv}

\usepackage[utf8]{inputenc} 
\usepackage[T1]{fontenc}    
\usepackage{hyperref}       
\usepackage{url}            
\usepackage{booktabs}       
\usepackage{amsfonts}       
\usepackage{nicefrac}       
\usepackage{microtype}      
\usepackage{lipsum}

\usepackage{graphicx}
\usepackage{todonotes}
\usepackage{fancyref}
\usepackage{caption}
\usepackage{subcaption}

\title{Tendrils of Crime:\\ Visualizing the Diffusion of Stolen Bitcoins}

\author{
  Mansoor Ahmed-Rengers\\
  Computer Laboratory\\
  University of Cambridge\\
  \And 
  Ilia Shumailov\\
  Computer Laboratory\\
  University of Cambridge\\
  \And
  Ross Anderson\\
  Computer Laboratory\\
  University of Cambridge\\
}

\begin{document}
\maketitle

\begin{abstract}
The first six months of 2018 have seen cryptocurrency thefts of \$761 million, and the technology is also the latest and greatest tool for money laundering.  This increase in crime has caused both researchers and law enforcement to look for ways to trace criminal proceeds. Although tracing algorithms have improved recently, they still yield an enormous amount of data of which very few datapoints are relevant or interesting to investigators, let alone ordinary bitcoin owners interested in provenance. In this work we describe efforts to visualize relevant data on a blockchain. To accomplish this we come up with a graphical model to represent the stolen coins and then implement this using a variety of visualization techniques.
\end{abstract}


\section{Introduction}

All Bitcoin transactions are written on the blockchain, a public append-only file. Tracing transactions might seem trivial, given the linear nature of the data structure. And there are already many visualizations of Bitcoin, ranging from simple diagrams of the transactions within each block to more involved projects showing clusters of communities within the network\cite{coin_viz,Battista_2015}. However, things are not so simple when one tries to analyse provenance information such as the flow of stolen coins.

We need to first understand the context of this research. The next section will provide some background on how Bitcoin transactions work. Next, we will look into what taint tracking is and why it is required; after that we will look at tracking techniques in the existing literature and why we chose one particular method. Next we introduce the difficulties with visualizing this tracking data and then present our solutions. We finally discuss the related work and conclude by pointing at avenues for future research.

\section{Bitcoin Primer}
In the interest of brevity, we abstract and simplify some of the relevant features of Bitcoin transactions. For a more thorough explanation, we direct the reader to the original paper~\cite{Nakamoto_bitcoin:a} or to the standard textbook~\cite{narayanan2016bitcoin}.

\subsection{Transactions}
To perform a Bitcoin transaction, you must first locate an Unspent Transaction Output (UTXO) for which you have a signing key, and spend it by signing it over to someone else. Essentially, the total amount of bitcoin you can spend is the total amount of UTXOs attributed to public keys whose private keys are in your control.

More generally, each transaction in Bitcoin is a signed blob that is interpreted by Bitcoin’s scripting system, called ``Script''. Each valid transaction consists of a set of input UTXOs, a set of signatures that verify using the public keys associated with those UTXOs, a set of output addresses, and an amount of cryptocurrency to be sent to each of the outputs. 

It is impossible to subdivide a UTXO, so if Bob wants to pay Alice 0.5 bitcoins but his savings are in the form of a single UTXO worth 50 bitcoins, then he has to make a transaction with two outputs: one to Alice (for 0.5 bitcoins), and one to a change address owned by himself (for 49.5 bitcoins). As a result, many bitcoin transactions have multiple outputs, and public keys in bitcoin tend to be short-lived. It is standard practice for a wallet application to generate a new keypair for each transaction and use the public key as the change address. 

Transactions can refer to UTXOs in blocks of many different ages. So while the first input to a transaction could be from the block immediately preceding the current one, the second could be from a block two years ago (roughly 50,000 blocks). Such hop lengths make temporal visualizations of bitcoin transactions quite problematic.

\subsection{A Loose Transaction Taxonomy}
For our purposes, we classify bitcoin transactions into the following types:

\begin{description}
  \item[1-to-1 transactions] \hfill \\ Transactions where a single UTXO is sent to a single output. These are often used as building blocks in more complex payment schemes.
  \item[Many-to-2 transactions] \hfill \\ The workhorse of bitcoin transactions; as discussed, these are a natural consequence of the indivisibility of UTXOs, and most legitimate transactions belong in this category.
  \item[1-to-many transaction] \hfill \\ These are quite rare since normal payments to multiple entities are executed by most wallets as a chain of transactions. 1-to-many transactions are often used in money-laundering schemes (also known as mixes) to split crime proceeds proceeds into many wallets in order to make tracing difficult.
  \item[Many-to-many transactions] \hfill \\ These are like 1-to-many transactions except that they have multiple input UTXOs. They are the second component in a typical mix; they shuffle cryptocurrency between different keys, mostly controlled by the same people.
\end{description}

Where transactions have many inputs and the inputs are signed by different keys, this provides extra information to the analyst, namely that the keys in question were under the same control. Heuristics like this enable analysts to cluster related transactions~\cite{Meiklejohn:2013:FBC:2504730.2504747}.

\section{Taint Tracking}
A specialist analysis firm has reported that in the first six months of 2018, 761 million dollars’ worth of cryptocurrencies have been stolen~\cite{ciphertrace}. Even if we only count major reported thefts from exchanges, perhaps 6-9\% of the bitcoins in circulation have been stolen at least once~\cite{Lee2017}; the true number is undoubtedly higher. If one is to make good the victims of these crimes, then we need to be able to track down stolen or otherwise tainted bitcoins. 

Bitcoin tracing is also important for law enforcement officers, regulators and researchers investigating ransomware, sanctions busting, online drug trafficking and other crimes facilitated by cryptocurrency. And the legal status of a bitcoin UTXO may depend on its history. Referring to figures 1--3, the red taint might mean that a bitcoin was stolen, green that it passed through the hands of an Iranian company under international sanctions, blue that it passed through a mix in contravention of money-laundering regulations and yellow that it was used to buy and sell drugs on AlphaBay. In the first case, it will still normally belong to the theft victim, who could sue to recover it. In the second, third and fourth, its owner may be prosecuted under applicable law. In the second, an owner who was a banker might be at risk of losing their licence. In the fourth, it may also be liable to particularly stringent asset-forfeiture laws; any wallet containing drug proceeds may be seized in many jurisdictions, with the onus then falling on the owner to prove honest provenance of any sums they wish to recover. 

\subsection{Status Quo}

For a while, Bitcoin researchers focused on two ways of doing tracing: poison and haircut. To illustrate the difference, suppose you have a wallet with three stolen bitcoin and seven freshly-mined ones. Then under poison all the coins you spend from this wallet are considered 100\% stolen, while under haircut they are reckoned to be 30\% stolen.

\begin{figure}[t]
  \centering
  \begin{subfigure}[b]{0.35\textwidth}
  	\includegraphics[width=\linewidth]{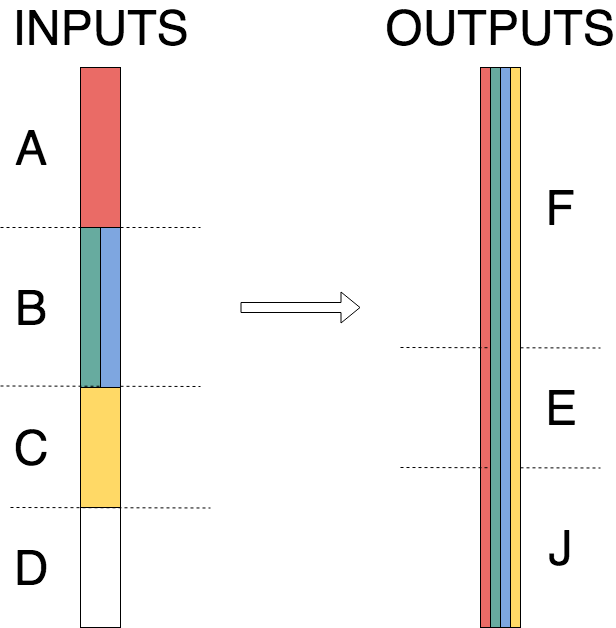}
    \caption{Poison Tainting}
	\label{fig:poison}
  \end{subfigure}
  \hspace{2cm}
  \begin{subfigure}[b]{0.35\textwidth}
  	\includegraphics[width=\linewidth]{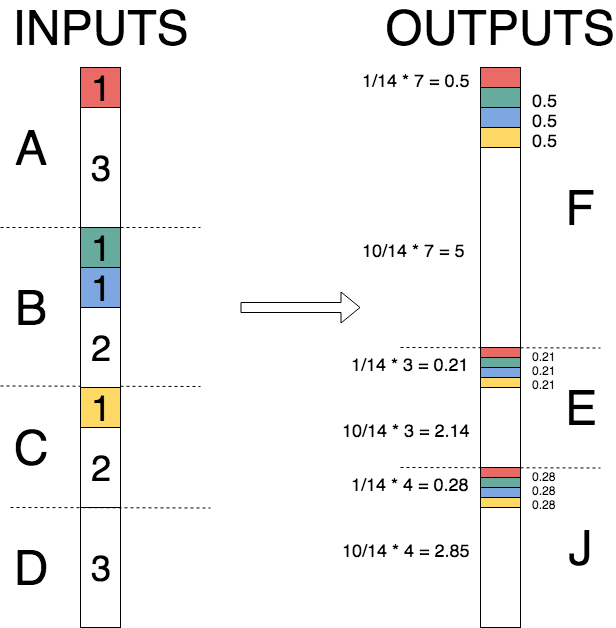}
    \caption{Haircut Tainting}
	\label{fig:haircut}
  \end{subfigure}
  \caption{The long-standing methods proposed within the Bitcoin community for taint tracking. }
\end{figure}

This goes across to multiple types of taint. In poison, if you have inputs with four different kinds of taint then all the outputs are tainted with everything. This leads to rapid taint contagion. Figure~\ref{fig:poison} illustrates poison tainting.

Haircut is only slightly different. Here, taint is not binary but fractional. So, instead of saying that all the outputs are tainted with the four kinds of taint, we associate a fractional value to the taint. If half of the input was tainted red then all the outputs are half red-tainted. Taint diffuses quickly through the network as in poison, but the result is rapid taint diffusion, rather than contagion. Figure~\ref{fig:haircut} illustrates haircut tainting.

To put numbers to the diffusion, we ran poison and haircut on a couple of major thefts from 2014 and found that by 2017 more than 90\% of wallets active on the network were tainted. This diffusion prevents any sensible recourse for victims -- if we were to recover the 9\% of stolen bitcoin and refund the victims, we might as well levy a 9\% tax on all users. That is politically and technically impractical. What we need is a deterministic manner of tainting that does not diffuse wildly.

\subsection{FIFO Taint Tracking}

\begin{center}
	  \begin{figure}
      \centering
  	\includegraphics[width=0.35\linewidth]{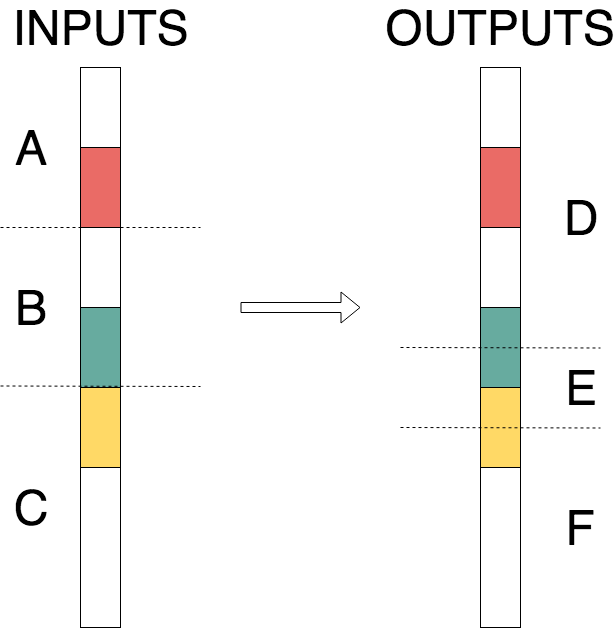}
    \caption{FIFO Tainting.}
	\label{fig:fifo}
  \end{figure}
	\end{center}

The diffusion problem is tackled by recent work by Anderson et al.~\cite{taint,taintredux}. They proceed from on Clayton's case -- a legal precedent in London in 1816 and in force throughout the UK, Canada and many other Commonwealth countries. The judge in that case decided that funds whose ownership is under dispute must be tracked through accounts on a strict First-In-First-Out (FIFO) basis. A natural conclusion is that taint in a cryptocurrency should be tracked in this way. This greatly cuts the diffusion as taint is conserved. It is shown in Figure~\ref{fig:fifo}.

Each bitcoin is divided into 100 million satoshi, and each satoshi is unique, in that it has a unique and public history. The data to enable tracing is built into the system; we just need the right algorithm to parse it; and FIFO appears to be that algorithm.

The FIFO principle is well-known in computer science as well as in law. FIFO tracking of disputed cryptocurrency turns out to be lossless and deterministically reversible. In addition to tracking a stolen bitcoin forwards -- as one has to do with the poison or haircut methods -- one can track a current UTXO backwards to all the reward blocks in which its component satoshi came into existence. This also makes for a much cleaner implementation. The tricky bit is the handling of transaction fees but once that's done, we can track the provenance of any satoshi. 

\subsection{Taintchain}

We implemented FIFO tracking and built it into a system we call the taintchain. This starts off from a set of reported thefts or other crimes and propagates the taint backwards or forwards throughout the entire blockchain. If working forwards, start from all tainted transaction outputs and mark all the affected satoshis as tainted until you reach the end of the blockchain. If working backwards, trace each UTXO of interest backwards and if at any point you encounter a taint, then return taint for the affected satoshis. This was described in~\cite{taint}.

The visualization problem we tackle is how to analyse the data generated by the taintchain system\footnote{Accessible at: \url{https://github.com/TaintChain}}.

\section{Visualizing Taint}
When we started analysing the taintchain, we ran into a number of issues. First is Big Data: just with 56 kinds of taint, we ended up with a dataset of about 450 GB. This grows about linearly as the user starts considering more crimes or more kinds of taints.

The second problem is that the things we’re looking for -- side effects of crime -- are not always amenable to algorithmic analysis. Different criminals use different strategies to lauder their money; and mixes are designed to be difficult to deal with.

We surmised that a good visual representation of the data might help us to spot patterns. Moreover, it would possibly make the taintchain more usable -- you could just enter your txhash and follow the taint. 

\subsection{Preliminary Model}
Our first prototype used a simple graphical model for our taintchain data. We represented each transaction as a vertex and each hop as an edge. By hop, we refer to the output of a transaction that has been used as an input somewhere else. Then we looked to represent our graph sensibly on-screen.

\begin{figure}
  \includegraphics[width=\linewidth]{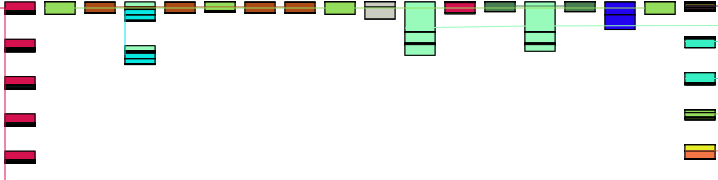}
  \caption{An illustrative image from our preliminary visualization showing multitaint movement.}\label{fig:multitaint}
\end{figure}

We decided to retain the chronological order and represent blocks as columns of transactions. Each transaction is a coloured rectangle where the colour reflects the kind of taint, and the size of rectangle reflects the number of satoshis tainted. Lastly, we decided to ignore clean satoshis as the data was sparse and required too much scrolling. We displayed this model as a static SVG graphic with click-to-reveal txhashes. Figure~\ref{fig:multitaint} shows an example.

\begin{figure}[!htb]
\minipage{0.49\textwidth}
  \includegraphics[width=\linewidth]{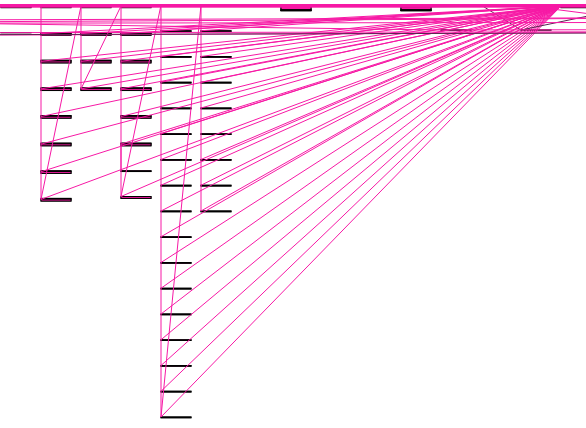}
  \caption{A collection pattern}\label{fig:coll}
\endminipage\hfill
\minipage{0.49\textwidth}
  \includegraphics[width=\linewidth]{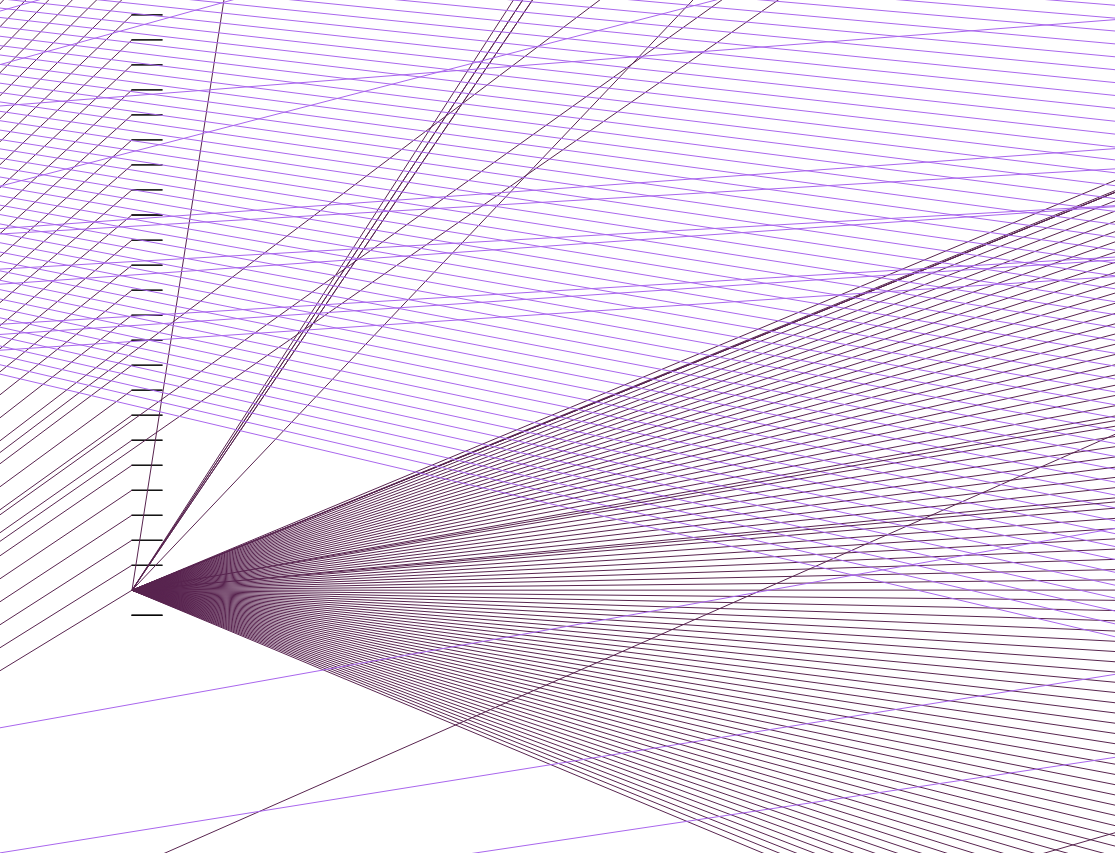}
  \caption{A splitting pattern}\label{fig:dist}
\endminipage
\end{figure}

To our surprise, even this rudimentary model gave us good results.  We were able to spot quite a few interesting patterns via the visualization that we wouldn’t have been able to see otherwise. For example, Figure~\ref{fig:coll} shows someone collecting crime proceeds, that they had initially split to many addresses, into a single address. We call this a collection pattern and we observed similar patterns many times; in some of the instances, we were able to connect the collection address to illegal gambling sites.

Figure~\ref{fig:dist} shows the converse of a collection pattern: a splitting pattern. These may occur close to the time of a crime as criminals try to cover their tracks by feeding their loot into systems that divide their winnings into hundreds of tiny transactions.

\subsection{Limitations of Preliminary Model}

\begin{figure}[!htb]
\minipage{0.49\textwidth}
  \includegraphics[width=\linewidth]{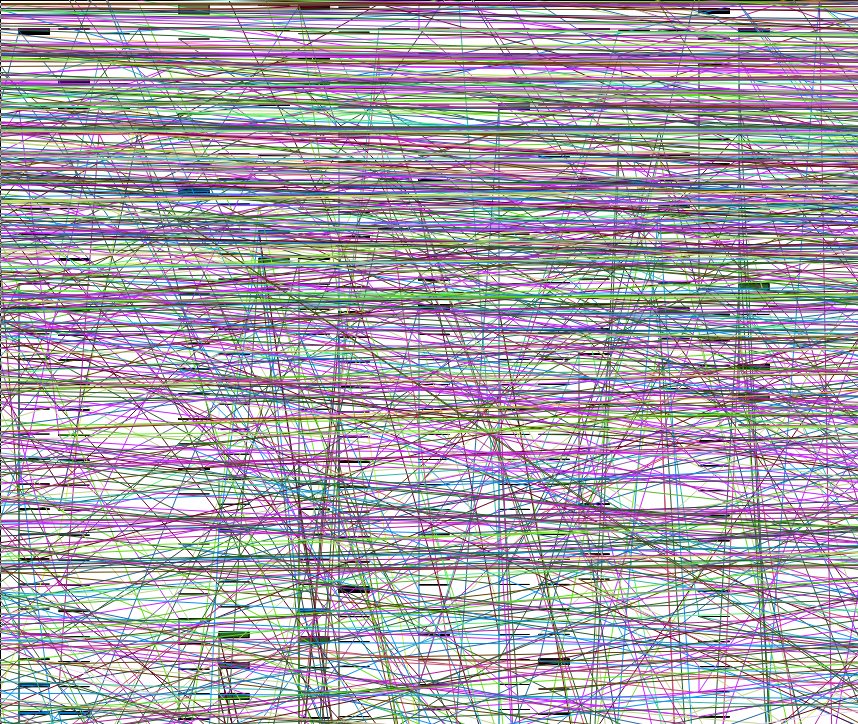}
  \caption{\textbf{Transaction density.} The sheer number of tainted transactions renders some sections of the taintgraph uninterpretable.}\label{fig:uninterpr}
\endminipage\hfill
\minipage{0.49\textwidth}
  \includegraphics[width=\linewidth]{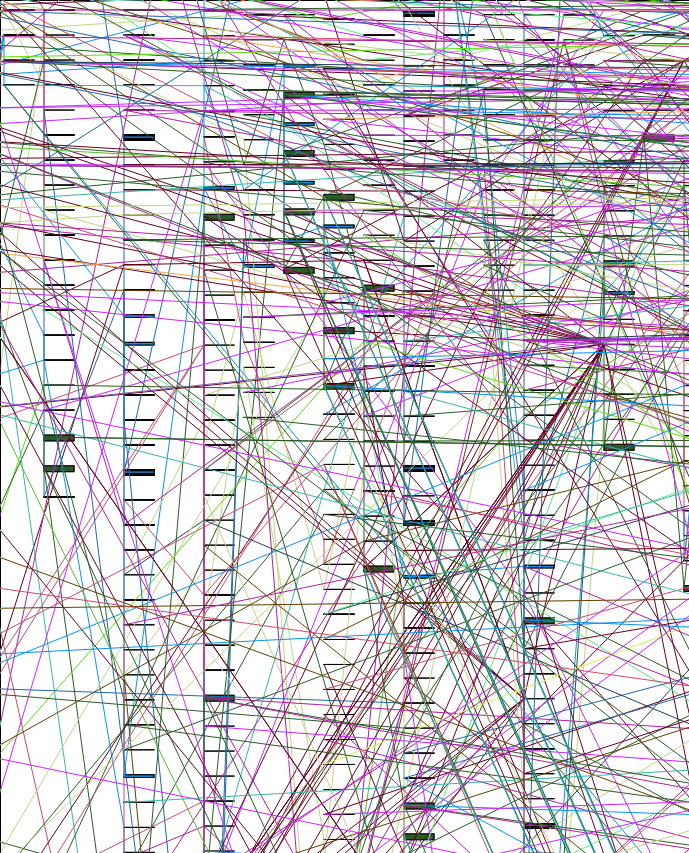}
  \caption{\textbf{Complex collection pattern.} We can see here the attempts by various actors to collect funds. However, this is difficult to spot due to the high degree of collocation of transactions.}\label{fig:collection_complex}
\endminipage
\end{figure}

One of the main problems we faced was sheer data density. In Figure~\ref{fig:uninterpr} we are displaying only four kinds of taint and yet it is strenuous to follow the many lines. Increased spacing is not a solution here as that would result in an unmanageable amount of vertical scrolling.

Another problem we faced was that taint tends to overlap, as shown in Figure~\ref{fig:collection_complex}. In that case, do we retain just one colour? Or do we create a new colour to represent the combination? 

\subsection{Interactive Visualization}

We therefore decided to rethink our approach. The second prototype makes the graph interactive so the user can choose which information is relevant to her on the fly. Secondly, we decided to make the edges more meaningful. Rather than just show a connection between nodes, we incorporated the proportion of satoshis transferred in each hop into the edges. Lastly, we decided to abandon displaying the blocks as columns of transactions; instead we now focussed solely on the transaction flows and included the block information as a hint box displayed on mouse hover. Thus, now the depth of a vertex does not necessarily relate to its chronological order.

\begin{figure}[t]
  \centering
  \begin{subfigure}[b]{0.7\textwidth}
  	\includegraphics[width=\linewidth]{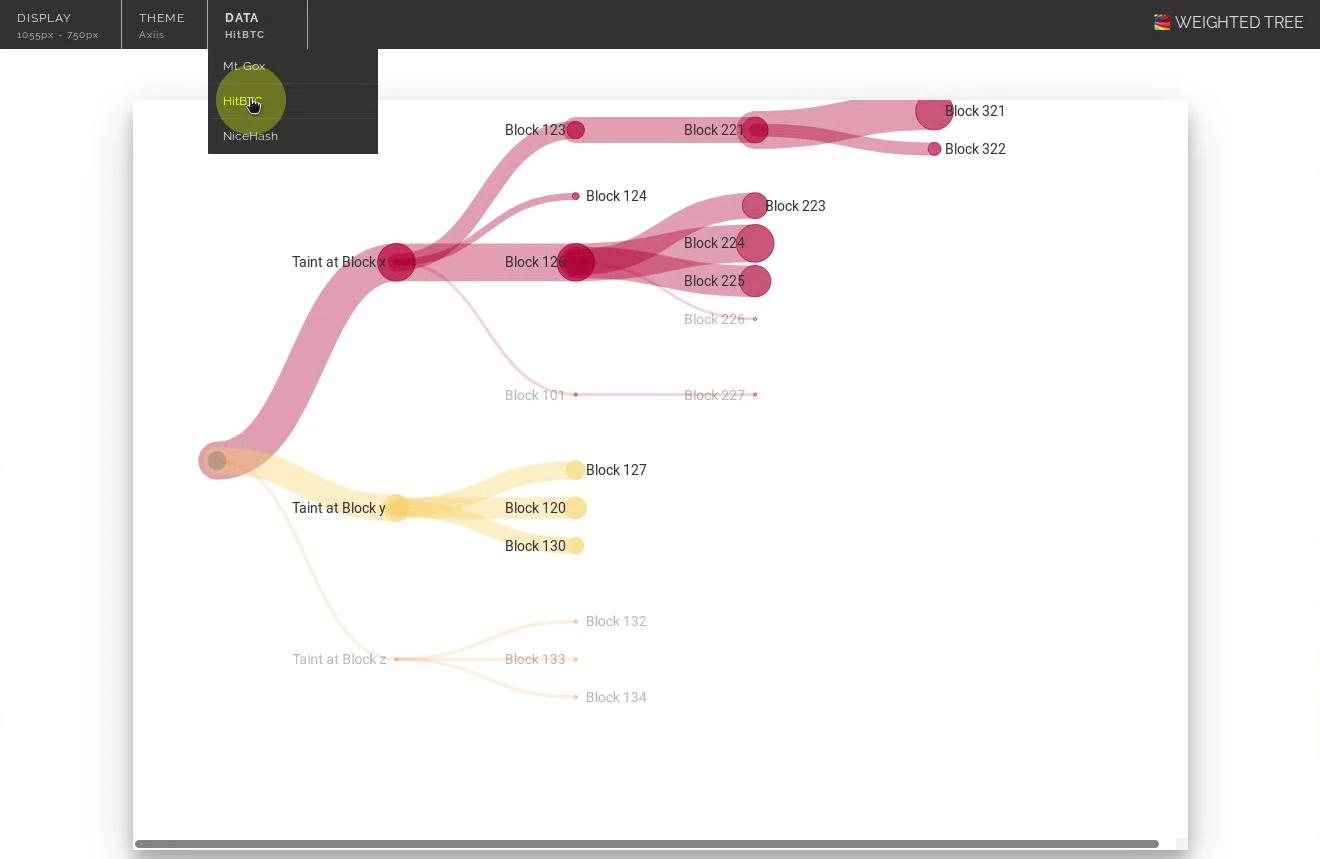}

  \end{subfigure}
  \hspace{2cm}
  \begin{subfigure}[b]{0.7\textwidth}
  	\includegraphics[width=\linewidth]{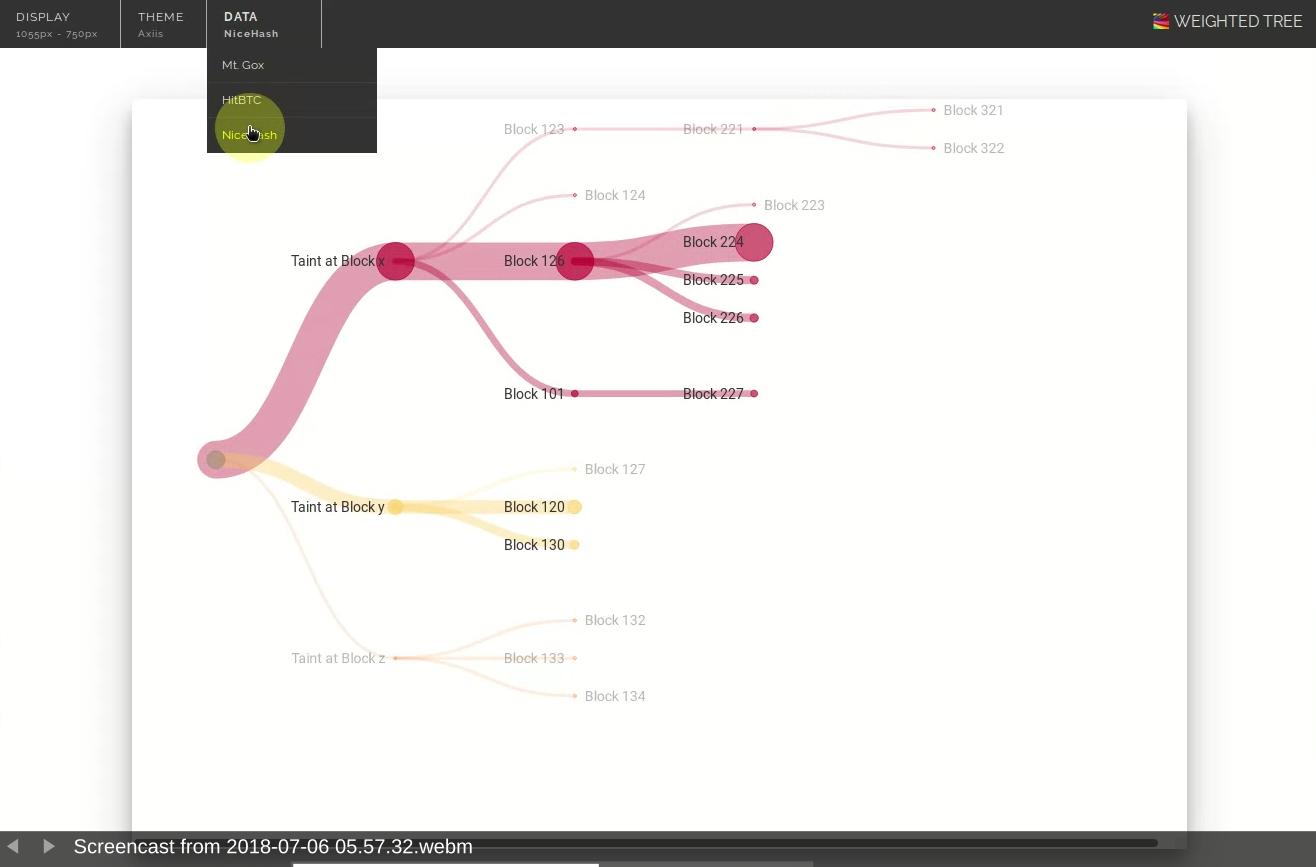}

  \end{subfigure}
  \caption{These screenshots illustrate how the graph dynamically changes based on the taint type currently selected.}
  \label{fig:overlapdynamic}
\end{figure}

One of the problems that immediately vanished by the move to interactive representation was that of taint overlap. In our new system, we simply included a drop-down menu where the user can choose the taint type of interest and the graph adjusts its edges accordingly. Figure~\ref{fig:overlapdynamic} shows this in action.

Making the graph interactive came at a cost, though, since now we want to store as much of the taintgraph in RAM instead of on disk for greater responsiveness. Second, since the graph expands on click, random exploration could lead to many uninteresting paths being followed.

\begin{center}
	  \begin{figure}
      \centering
  	\includegraphics[width=0.75\linewidth]{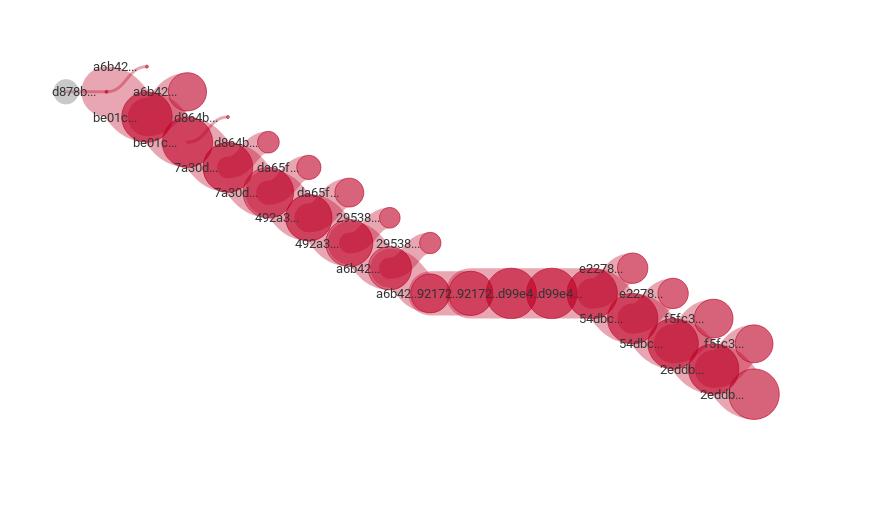}
    \caption{A peeling chain, discovered by following the larger branch at each vertex.}
	\label{fig:peel}
  \end{figure}
	\end{center}

We discovered some interesting patterns using this visualization. We were able to find multiple instances of \emph{peeling chains}, as shown in Figure~\ref{fig:peel}. These are usually seen used by exchanges or gambling sites -- in this case a notorious criminal exchange. Its operators would pool their money into a single wallet and then they would pay their customers successively, each time sending most of it to themselves at a change address. In this case, we can also see that this criminal exchange tried to hide their identity by shuffling their keys four times.

\begin{center}
	  \begin{figure}
      \centering
  	\includegraphics[width=0.5\linewidth]{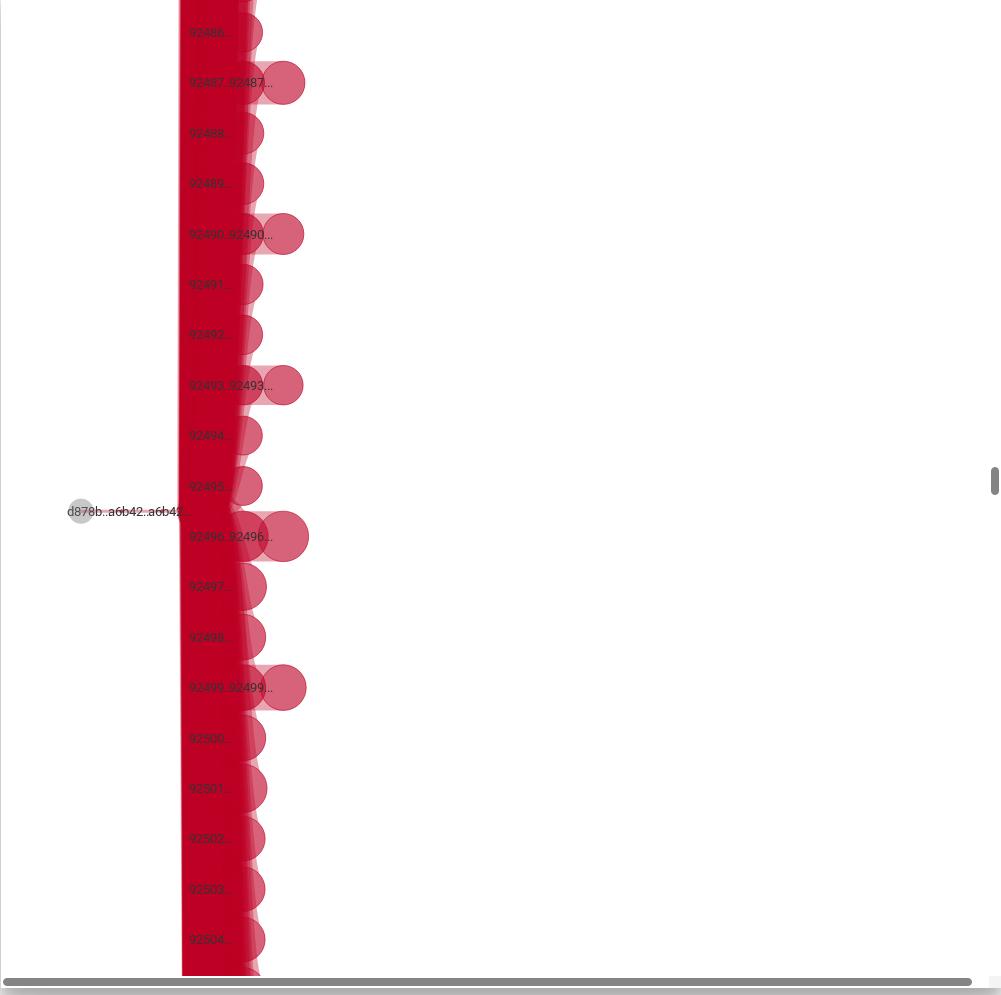}
    \caption{Exhaustive vertical scrolling due to high outdegrees of transactions. Notice the scroll bar on the right.}
	\label{fig:vertical}
  \end{figure}
	\end{center}

However, although these visualisations are better than nothing, there still remains much to be done. A fundamental issue seems to be that of the large outdegree of some transactions. A transaction can have an (effectively) unbounded number of outputs, which makes visualizations difficult. Figure~\ref{fig:vertical} illustrates this difficulty. One possible solution is to have a filter for transactions: collapse all the outputs below a certain threshold. This would give a cleaner display image, but might hamper investigations. We are still exploring effective aggregations that do not result in egregious information loss.

\section{Related work}

A number of previous attempts have been made to visualize the Bitcoin network, with most of them focusing on some specific task. Early attempts were concerned with simple property representations e.g. Reid and Harrigan featured loglog plots of graph centrality measurements, graph representations with sizes of nodes showing the amounts of money transferred, geographical activity acquired through IP address mappings from Bitcoin Faucet, and graph representation of poison tainting~\cite{Reid_harrigan_2011}. 

Later came systems like BitIodine with graph-like outputs to support commonly available graph representation tools~\cite{Spagnuolo_2014}. Graph approaches to transaction visualization were also adopted for educational purposes by systems like CoinVis~\cite{coin_viz}, while bitcoin-tx-graph-visualizer used alluvial diagrams to show Bitcoin movement~\cite{tx_graph_vis}. 

A more mature system was BitConeView, presented by Battista and Donato in 2015~\cite{Battista_2015}. This was among the first to provide a sensible GUI to inspect how a particular UTXO propagated through the network. In order to explain what it means for money to move, the authors came up with `purity' -- basically a version of haircut tainting. They only evaluated the usability of their system informally, and came to the conclusion that more improvements were necessary to the way purity was presented to the user.

McGinn et al. devised a graph visualization of blockchain that allowed them to detect laundering activity and several denial-of-service attacks~\cite{Dan_2016}. Unlike previous approaches, they made use of top-down system-wide visualization to understand transaction patterns. The follow-up paper from Molina et al. proposed an extension to a global view, in which graph analysis is aided by human intuition~\cite{MOLINASOLANA2017227}. 

In our system we set out to learn from and build on all of this previous work. In particular, we focus on data representation in taint propagation when a taint graph becomes too massive for humans to comprehend.

Unlike BitConduit and similar systems, we are not doing any actor characterization in our visualisation tool~\cite{kinkeldeyhal_2017}. The generation of graph colours is exogenous, relying on external theft reports or of software that analyses patterns of mixes, ransomware and other undesirable activity.

\section{Future Work and Conclusion}

In this short paper, we have presented a system for visualizing FIFO taint diffusion without any information-losing abstractions. This system has helped us spot interesting patterns that hint at the operational techniques of criminals operating on the Bitcoin network. We have made this system publicly available for anyone to use and modify.

It still suffers from a number of shortcomings that invite further work. One avenue for research would be to explore different heuristics to portray the data more concisely. One might aim at a system that presents a global, zoomed-out view of the data and successively introduces more information as the user explores a particular pattern on the blockchain. Another direction would be to highlight suspicious patterns of transactions automatically, for example, by marking coins that have recently emerged from a flurry of splits and merges. There are many other plausible heuristics to explore, a lot of data to analyse, and real social problems to tackle.

\bibliographystyle{unsrt}  
\bibliography{sample}

\end{document}